\documentclass[aps,prb,reprint,groupedaddress]{revtex4-2}

\usepackage[dvipdfmx]{graphicx}
\usepackage{dcolumn}
\usepackage{bm}
\usepackage{amsmath, amssymb}
\usepackage{upgreek}
\usepackage{here}

\newcommand{\rxy}{\rho_{xy}}

\newcommand{\rxx}{\rho_{xx}}
\newcommand{\muea}{{\mu}_{\rm e1}}
\newcommand{\mueb}{{\mu}_{\rm e2}}
\newcommand{\mue}{{\mu}_{\rm e}}
\newcommand{\muh}{{\mu}_{\rm h}}
\newcommand{\nea}{n_{\rm e1}}
\newcommand{\neb}{n_{\rm e2}}
\newcommand{\nd}{n_{\rm e}}
\newcommand{\nh}{n_{\rm h}}
\newcommand{\Tc}{T_{\rm c}}
\newcommand{\Ts}{T_{\rm s}}
\newcommand{\Tczero}{$T_{\rm c}^{\rm zero}$}
\newcommand{\Tconset}{$T_{\rm c}^{\rm onset}$}

\begin{document}

\title{Control of band structure of FeSe single crystals via biaxial strain}

\author{M. Nakajima}%
\email[]{nakajima@phys.sci.osaka-u.ac.jp}

\author{Y. Ohata}
\author{S. Tajima}
\affiliation{Department of Physics, Osaka University, Osaka 560-0043, Japan}


\begin{abstract}
We performed systematic transport measurements on FeSe single crystals with applying in-plane biaxial strain $\varepsilon$ ranging from $-0.96\%$ to 0.23{\%}. Biaxial strain was introduced by firmly gluing samples to various substrate materials with different thermal expansion. With increasing ${\varepsilon}$, structural and superconducting transition temperatures monotonically increased and decreased, respectively. We analyzed magneto-transport results using a compensated three-carrier model. The evaluated densities of hole and electron carriers systematically changed with strain. This indicates that we succeeded in controlling the band structure of single-crystalline FeSe.
\end{abstract}

\maketitle

\section{Introduction}

Application of strain is one of the effective ways to engineer the band structure of materials. We can tune the properties of materials by strain through the control of the band structure. For example, strained silicon technology has been successful in improving device performance~\cite{Strain-Engineering-Tsutsui}. Recently, there is a growing interest in strain engineering in two-dimensional materials, such as graphene and transition metal dichalcogenides, the splendid properties of which are considered to be useful for a wide range of applications~\cite{Strain-Engineering-Sun}. On the other hand, strain has been underused for bulk single crystals because of the difficulty in applying homogeneous strain. Recent development of piezoelectric-based pressure cells~\cite{Hicks-Sr2RuO4} boosted measurements on uniaxially strained single crystals, but there are few studies using biaxial strain.

Iron-based superconductors are suitable materials to study the strain dependence of physical properties because their band structures are sensitive to the local crystal structure, in particular, the shape of tetrahedrons composed of iron and pnictogen/chalcogen atoms. Biaxial strain deforms the crystal lattice without introducing disorder. This feature is not shared by elemental substitution with isovalent atoms giving rise to a chemical-pressure effect.

Among iron-based superconductors, FeSe serves as an intriguing case of the interplay between superconductivity, magnetism, and electronic nematicity, which can be tuned by chemical substitution and application of physical pressure~\cite{Bohmer2018,Shibauchi2020}. FeSe exhibits a tetragonal-to-orthorhombic structural phase trainsition at $\Ts \sim$ 90 K and is superconducting below $\sim$ 8 K. A prominent feature of FeSe is its very small effective Fermi energy~\cite{Kasahara-Crossover}, and thus a change in the crystal structure is expected to make a large impact on physical properties. Thin films of FeSe can be grown on various substrates, leading to various degrees of in-plane strain~\cite{Nabeshima-Film-JJAP}. Angle-resolved photoemission spectroscopy (ARPES) on single crystals and thin films of FeSe revealed a systematic variation of the band structure with strain~\cite{Phan-ARPES}. Reflecting the change in the band structure, a superconducting transition temperature ${\Tc}$ shows a systematic strain dependence~\cite{Nabeshima-Film-JJAP}. Although FeSe thin films definitely offer an opportunity to study a variation of physical properties with biaxial strain, the sample quality varies depending on substrate materials. This weakness can be overcome by studies on biaxially strained single crystals.

In this study, we demonstrate that the band structure of FeSe single crystals was systematically controlled by applying biaxial strain, evidenced by electronic transport measurements. FeSe single crystals were affixed on substrates, resulting in the deformation of FeSe with substrates when cooling down. There are a few studies in which biaxial strain was induced by such a method~\cite{Wang-ST-1,Bohmer-Ca122,Fente2018}, but no systematic study has been reported so far. In the present work, we used several substrate materials and studied a systematic strain dependence of physical properties of FeSe. It turned out that tensite biaxial strain gives rise to an increase in $\Ts$. We also observed an increase (decrease) in ${\Tc}$ with compressive (tensile) strain, which follows the result for thin films~\cite{Nabeshima-Film-JJAP}. The analysis of magneto-transport results using a compensated three-carrier model revealed that the densities of all the three carriers systematically change with strain. This indicates that we succeeded in controlling the band structure of single-crystalline FeSe.

\section{Experimental}

\newcommand{\Rabs}{${\rho}_{\rm 300K}=0.45 \hspace{1mm} {\rm m}{\Omega}{\hspace{0.5mm}}{\rm cm}$}

Single crystals of FeSe were grown by a chemical vapor transport method as described elsewhere~\cite{Bohmer-Variation, Nakajima-Lattice}. To suppress dispersion of sample quality as much as possible, single crystals measured in this study were taken from the same batch. In-plane biaxial strain was applied on single crystals attached on substrates by means of the difference of thermal-expansion coefficients between samples and substrates. Polycarbonate (C.I. Takiron Corporation, PS610), Stycast\textsuperscript{\textregistered} 2850FTJ epoxy (Henkel Ablestik Japan Ltd., with Catalyst 9), soda-lime glass (AS ONE Corporation, ASLAB Slide Glass), borosilicate glass (SCHOTT AG, TEMPAX Float\textsuperscript{\textregistered}), and quartz glass (TOSOH Quartz Corporation, Synthetic Silica Glass, ES grade) were utilized as substrates. FeSe samples were attached using cyanoacrylate adhesives (Tokyo Measuring Instruments Laboratory Co., Ltd., CN adhesives) and epoxy adhesives (ThreeBond Co., Ltd., TB2086M) for the former two and the latter three substrates, respectively. To determine the thermal expansion of the five substrates, we measured the temperature dependence of electrical resistance of a strain gauge attached on the substrates as well as a copper plate as a reference material~\cite{Cu-Thermal}. FeSe single crystals were cut into a rectangular shape, the typical size of which was 0.7--1.0 mm in length and 0.3--0.5 mm in width. The attached samples were cleaved to fully transmit the strain to them. The thickness of the sample, typically 5 ${\upmu}$m or less, was derived from the resistivity value of FeSe at $T=300$ K: {\Rabs} (See Appendix A). Electrical resistivity was measured using a standard four-probe method, and electrical contacts were made with gold paste. Hall effect and magnetoresistance were measured with magnetic fields up to 7 T applied parallel to the $c$ axis. Hall coefficients were determined in the zero-field limit.

\section{Results}

\begin{figure}
	\includegraphics[width=8cm]{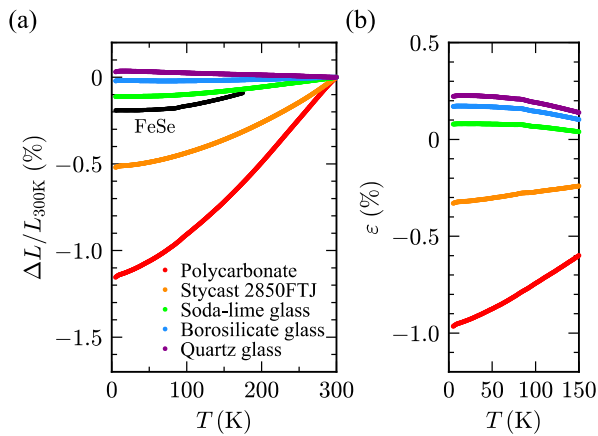}
	\caption{\label{fig:1}
		(a) Temperature dependence of the relative length changes $\Delta L/L_{\rm 300K}$ of polycarbonate, Stycast 2850FTJ, soda-lime glass, borosilicate glass, and quartz glass. The data for the in-plane length of twinned FeSe are also plotted~\cite{Bohmer-Lack}. (b) Temperature dependence of in-plane strain imposed on FeSe single crystals on the five substrates.}
\end{figure}

Figure~\ref{fig:1}(a) shows the relative length changes ${\Delta}L/L_{\rm 300 K}$ of the five substrate materials as a function of temperature. We also show the data for the in-plane length of twinned FeSe~\cite{Bohmer-Lack}. Thermal contraction of FeSe is smaller than that of polycarbonate and Stycast 2850FTJ and larger than that of the three kinds of glasses. This indicates that compressive in-plane strain is imposed on FeSe attached on the polycarbonate and Stycast 2850FTJ substrates, while the glass substrates create tensile strain. Assuming that the length of FeSe exactly follows that of the substrate, we estimated the value of in-plane strain ${\varepsilon}$ transmitted to the sample. As shown in Fig.~\ref{fig:1}(b), the magnitude of the biaxial strain is monotonically increased with decreasing temperature. ${\varepsilon}$ varies with temperature, but it can be regarded as a constant for a limited temperature window. At 5 K, ${\varepsilon}$ reaches $-0.96\%$ and 0.23{\%} for the polycarbonate and the quartz-glass substrate, respectively.

\begin{figure}
	\includegraphics[width=8.4cm]{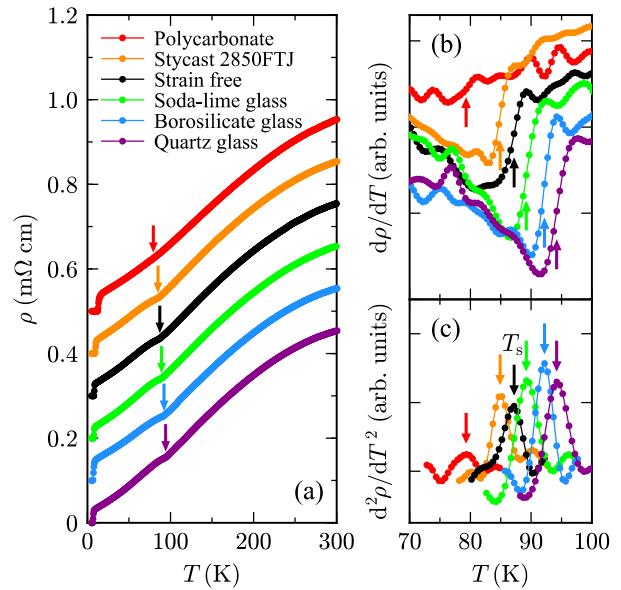}
	\caption{\label{fig:2}
		(a) Temperature dependence of resistivity ${\rho}(T)$ for the strain-free and strained FeSe. The curves are offset to avoid overlapping. (b) Temperature derivative and (c) second derivative of ${\rho}(T)$ in the magnified region around ${\Ts}$. Arrows indicate $\Ts$ determined from the peak position of d$^{2}\rho$/d$T^2$.}
\end{figure}

\begin{figure}
	\includegraphics[width=8cm]{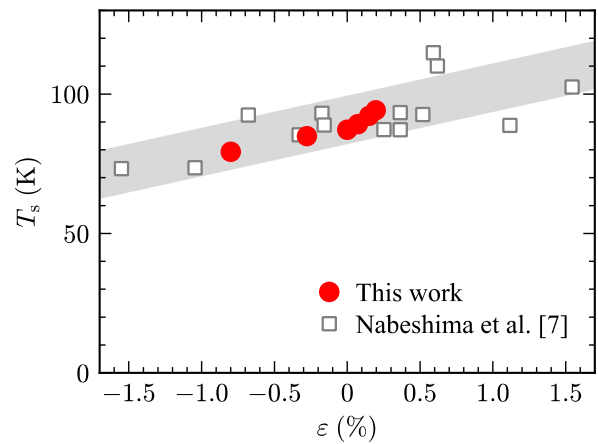}
	\caption{\label{fig:3}
		Strain dependence of ${\Ts}$ of FeSe. Square symbols represent ${\Ts}$ obtained from the study on thin films~\cite{Nabeshima-Film-JJAP}. The gray line is a guide to the eye.}
\end{figure}

In Fig.~\ref{fig:2}(a), we show the temperature dependence of resistivity ${\rho}(T)$ of FeSe on various substrates for a wide temperature range. A kink is observed at ${\Ts}$, which results in an abrupt change in the derivative [Fig.~\ref{fig:2}(b)]. For the polycarbonate substrate giving rise to most compressive strain, a kink structure is highly smeared. While B\"{o}hmer \textit{et al.}\ estimated $\Ts$ from the midpoint of the steplike anomaly of $\mathrm{d}\rho/\mathrm{d}T$~\cite{Bohmer-Variation}, we determined $\Ts$ from the peak position of the second derivative of ${\rho}(T)$, as indicated by arrows in Fig.~\ref{fig:2}(c). ${\Ts}$ is 87 K for the strain-free sample and systematically changes with strain. In Fig.~\ref{fig:3}, we plot ${\Ts}$ as a function of ${\varepsilon}$, the degree of the strain at ${\Ts}$. One can see a tendency that ${\Ts}$ increases with increasing ${\varepsilon}$, consistent with the thin-film study~\cite{Nabeshima-Film-JJAP}. This indicates that the present method works well to produce various degrees of biaxial strain.

\begin{figure}
	\includegraphics[width=8cm]{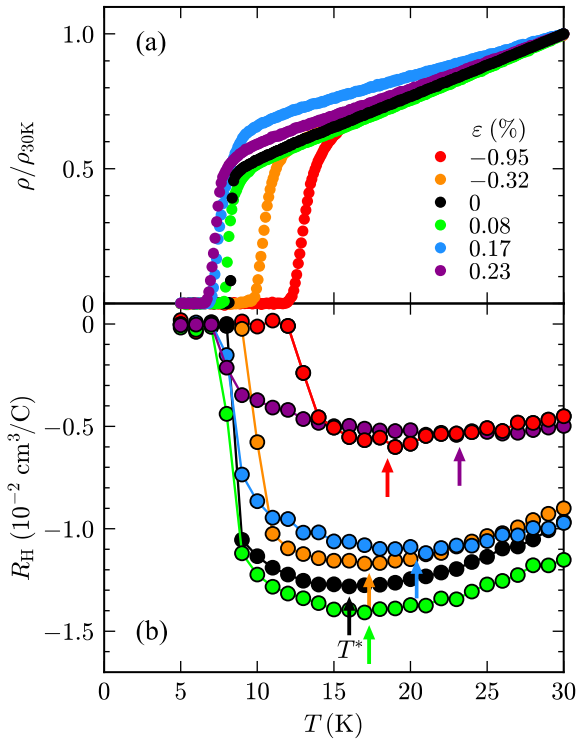}
	\caption{\label{fig:4}
		Temperature dependence of (a) electrical resistivity normalized at 30 K and (b) Hall coefficient for strain-free and strained FeSe. The displayed $\varepsilon$ corresponds to the value at $\Tc$. The arrows indicate the minima of $R_{\rm H}$ at $T^{*}$.}
\end{figure}

\begin{figure}
	\includegraphics[width=8cm]{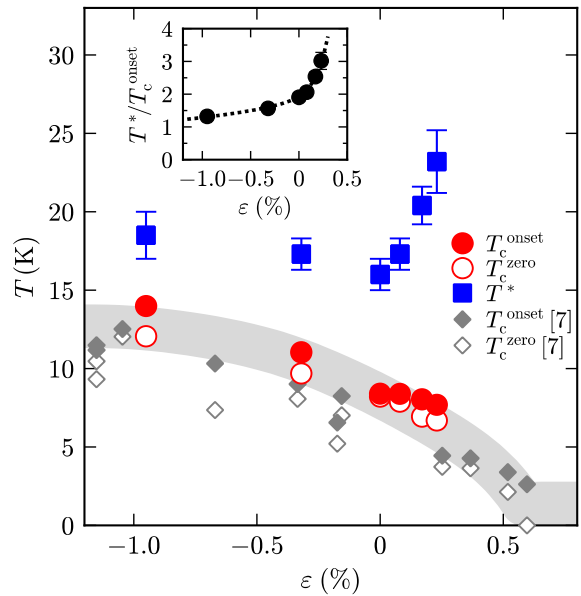}
	\caption{\label{fig:5}
		Strain dependence of {\Tczero}, {\Tconset}, and $T^{*}$. Diamond symbols represent ${\Tc}$ obtained from the study on thin films~\cite{Nabeshima-Film-JJAP}. The gray line is a guide to the eye. Inset shows the strain dependence of $T^{*}/$\Tconset{}.}
\end{figure}

Having confirmed the application of biaxial strain, we measured tranport properties at low temperatures. Figure~\ref{fig:4}(a) shows the temperature dependence of resistivity normalized to the value at 30 K for strain-free and strained FeSe. Although a slight broadening of the superconducting transition is seen for the strained samples, the systematic change in ${\Tc}$ with strain is clearly observed, indicating that the applied strain is satisfactorily homogeneous. The strain-free sample shows an abrupt drop of resistivity at 8.4 K and zero resistivity at 8.2 K. The onset ${\Tc}$ is enhanced with compressive strain to 14.0 K for ${\varepsilon}=-0.95\%$ and is decreased with tensile strain to 7.7 K for ${\varepsilon}=0.23${\%}. In Fig.~\ref{fig:5}, we plot $T_{\rm c}$ as a function of ${\varepsilon}$. The strain dependence of ${\Tc}$ is coincident with the study on thin films~\cite{Nabeshima-Film-JJAP}.

Figure~\ref{fig:4}(b) shows the temperature dependence of Hall coefficient $R_{\rm H}(T)$ below 30 K. The negative Hall coefficient indicates dominant contribution of electron carriers. In the strain-free sample, $R_{\rm H}(T)$ shows a minimum at $T^{*}=16$ K. This behavior is consistent with the previous study~\cite{Kasahara-Giant}, although the temperature in the present study is slightly lower. With applying compressive strain, $T^{*}$ is enhanced to 19 K for ${\varepsilon}=-0.95\%$ (Fig.~\ref{fig:5}). This seems to be linked with an increase in ${\Tc}$. In contrast, $T^{*}$ abruptly increases with tensile strain (23 K at $\varepsilon=0.23\%$), which is the opposite direction of the variation of ${\Tc}$. The origin of the minimum of $R_{\rm H}(T)$ at $T^{*}$ has yet been unclear, but a relation with superconducting fluctuations has been debated~\cite{Kasahara-Giant}. It is worth estimating $T^{*}/{\Tc}$ because the Hall effect can be a good probe to detect superconducting fluctuations~\cite{Hall-Fluctuation-2,Hall-Fluctuation-1}. As shown in the inset of Fig.~\ref{fig:5}, $T^{*}/{\Tc}$ shows a systematic increase with tensile strain, in contrast to the non-monotonic variation of $T^*$. This behavior might indicate that superconducting fluctuations can be controlled in terms of FeSe as a candidate superconductor in the crossover regime between the weak-coupling Bardeen-Cooper-Schrieffer and the strong-coupling Bose-Einstein-condensation limit~\cite{Kasahara-Crossover,Kasahara-Giant}.

\begin{figure}
	\includegraphics[width=8cm]{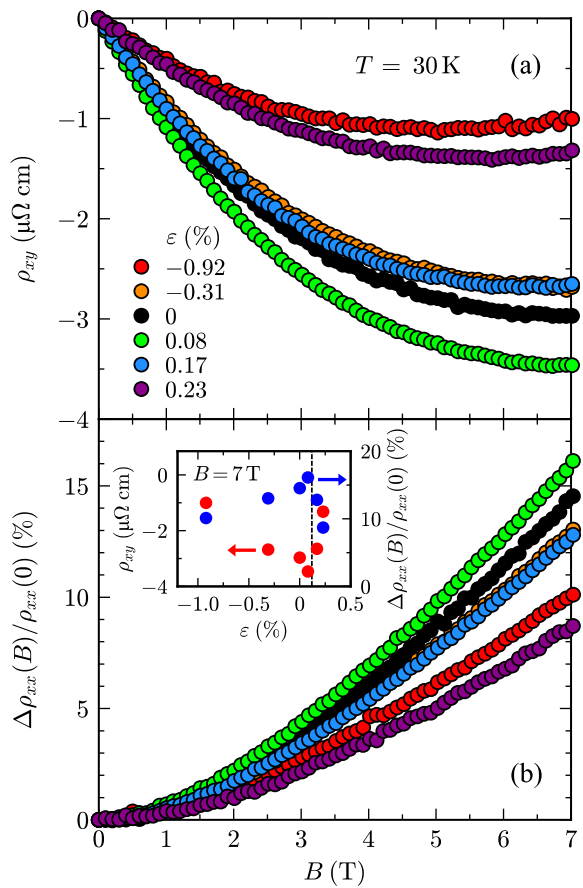}
	\caption{\label{fig:6}
		Magnetic-field dependence of (a) Hall resistivity $\rho_{xy}$ and (b) magnetoresistance $\Delta \rho_{xx}(B) / \rho_{xx}(0)$ of FeSe with various degrees of biaxial strain at 30 K. Inset shows the strain dependence of the Hall resistivity and the magnetoresistance at $B$ = 7 T. A vertical broken line separates distinct strain dependences, which implies the presence of a Lifshitz transition.}
\end{figure}

The strain-induced variation of ${\Tc}$ and $T^{*}$ infers a change in the band structure. To get firmer evidence, we conducted magneto-transport measurements. Figures~\ref{fig:6}(a) and \ref{fig:6}(b) show the magnetic-field dependence of Hall resistivity $\rho_{xy}$ and magnetoresistance $\Delta \rho_{xx}(B) / \rho_{xx}(0)$, respectively, up to 7 T for strain-free and strained FeSe at $T=30$ K. The absolute value of $\rho_{xy}$ increases with increasing $\varepsilon$ from $-0.92\%$ to $0.08\%$. $\rho_{xy}$ turns to decrease with further increasing $\varepsilon$. A similar behavior is seen for the magnetoresistance. With increasing $\varepsilon$, the magnetoresistance increases at first, takes the maximum at $\varepsilon=0.08\%$, and then decreases. The non-monotonic strain dependence is clearly seen in the inset of Fig.~\ref{fig:6}(b).

Since FeSe is a compensated metal, both holes and electrons contribute to the charge transport. The present result can be understood as a loss of electrons with high mobility exceeding $\varepsilon=0.08\%$. Relatively enhanced hole contribution makes $\rho_{xy}$ shift to the positive side. In addition, the reduction of mobility leads to the decrease in the magnetoresistance. These imply the drastic change in the Fermi surface with biaxial strain. Indeed, the change in the Fermi-surface topology is observed by ARPES on thin-film FeSe~\cite{Phan-ARPES}. The electron pocket at the Brillouin-zone corner, which is present for the strain-free sample, is absent for the film with tensile in-plane strain, resulting in tiny electron pockets away from the zone corner associated with the Dirac-cone-like band dispersion. Together with our magneto-transport result, the Lifshitz transition is likely to occur between $\varepsilon=0.08\%$ and $0.17\%$.

\section{Data analysis and discussion}

The observed nonlinear Hall effect indicates the presence of multiple carriers. In such a case, analysis with a two-carrier model has widely been performed. This model is, however, not appropriate to be applied to FeSe because a compensated two-carrier model with equal numbers of holes and electrons yields a linear Hall effect, in contrast to the present result shown in Fig.~\ref{fig:6}(a) (see Appendix B for more details).

\begin{figure*}
	\includegraphics[width=17.4cm]{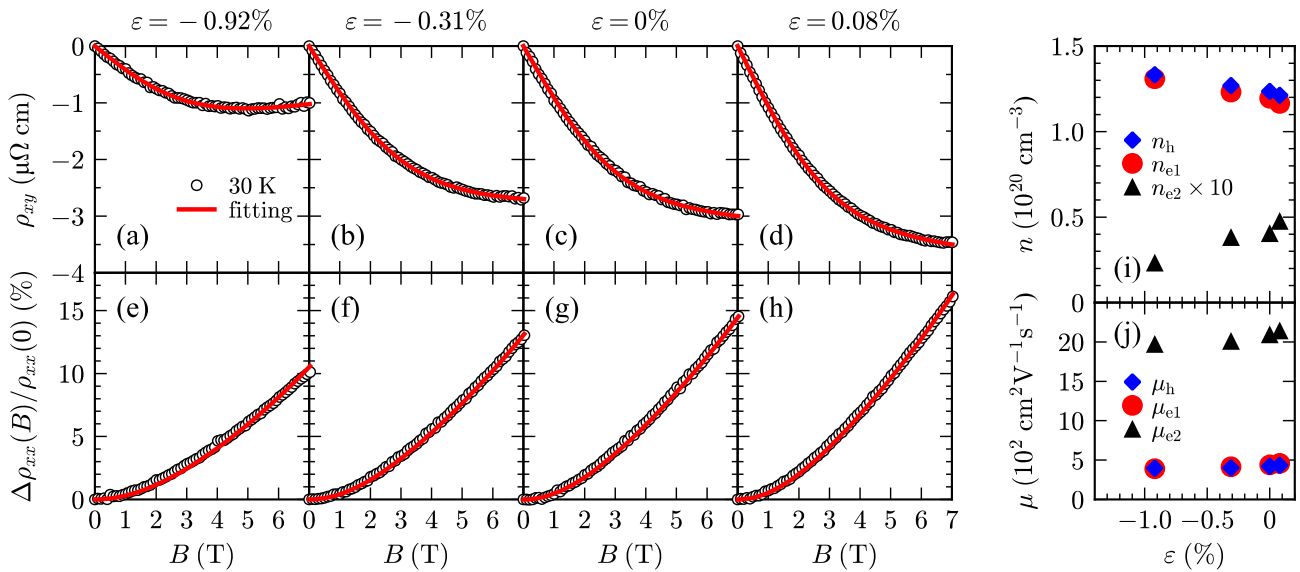}
	\caption{\label{fig:7}
		Magnetic-field dependence of (a)--(d) Hall resistivity and (e)--(h) magnetoresistance for strain-free and strained FeSe at 30 K. Fitting curves obtained using a compensated three-carrier model are also shown. Strain dependence of (i) carrier densities (${\nh}$, ${\nea}$, and ${\neb}$) and (j) mobilities (${\muh}$, ${\muea}$, and ${\mueb}$) extracted by the three-carrier fitting.}
\end{figure*}

\begingroup
\renewcommand{\arraystretch}{1.4}
\begin{table}
	\caption{\label{tab:T1}
		Carrier densities and mobilities extracted from the compensated three-carrier analysis for strain-free and strained FeSe at 30 K.}
	\begin{center}
		\begin{tabular}{p{7.3em}p{3em}p{4.5em}p{4.5em}p{4.5em}} \hline \hline
			Parameters & \hfil ${\varepsilon}$ (\%) \hfil & \hfil hole \hfil & \hfil electron 1 \hfil & \hfil electron 2 \hfil \\ \hline
			$n$ ($10^{20}$ cm$^{-3}$) & \hfil $-0.92$ \hfil & \hfil 1.332 \hfil & \hfil 1.309 \hfil & \hfil 0.023 \hfil  \\
			${\mu}$  (cm$^{2}$V$^{-1}$s$^{-1}$)       &                    & \hfil 396  \hfil & \hfil  390 \hfil & \hfil 1974 \hfil  \\ \hline
			$n$ ($10^{20}$ cm$^{-3}$) & \hfil $-0.31$ \hfil & \hfil 1.269 \hfil & \hfil 1.231 \hfil & \hfil 0.038 \hfil  \\
			${\mu}$  (cm$^{2}$V$^{-1}$s$^{-1}$)       &                    & \hfil 401  \hfil & \hfil  415 \hfil & \hfil 2014 \hfil  \\ \hline
			$n$ ($10^{20}$ cm$^{-3}$) & \hfil 0     \hfil & \hfil 1.236 \hfil & \hfil 1.195 \hfil & \hfil 0.041 \hfil  \\
			${\mu}$  (cm$^{2}$V$^{-1}$s$^{-1}$)       &                    & \hfil 420  \hfil & \hfil  439 \hfil & \hfil 2098 \hfil  \\ \hline
			$n$ ($10^{20}$ cm$^{-3}$) & \hfil 0.08 \hfil & \hfil 1.213 \hfil & \hfil 1.165 \hfil & \hfil 0.048 \hfil  \\
			${\mu}$  (cm$^{2}$V$^{-1}$s$^{-1}$)       &                    & \hfil 434  \hfil & \hfil  458 \hfil & \hfil 2146 \hfil  \\ \hline \hline
		\end{tabular}
	\end{center}
\end{table}
\endgroup

We thus analyzed the magneto-transport data using a three-carrier model~\cite{Kim1999,Ishida-Ba122} (See Appendix C for the formalism of this model). In light of the band structure of FeSe~\cite{ARPES-FeSe-1, ARPES-FeSe-2}, one can consider one hole and two electron bands contributing to the transport properties. Indeed, the previous transport studies pointed out that the three types of carriers well explain the experimental data~\cite{Watson2015}. The three-carrier model has six parameters: carrier densities ({$\nh$}, {$\nea$}, and {$\neb$}) and mobilities ({$\muh$}, {$\muea$}, and {$\mueb$}) for one hole and two electron carriers. Here, we put two constraints on this model. One is charge compensation required in the present system, i.e., ${\nh}={\nea}+{\neb}$. The other is that the calculated resistivity in zero field agrees with the value obtained experimentally [$1/{\rho}=e({\nh}{\muh}+{\nea}{\muea}+{\neb}{\mueb})$]. Using this compensated three-carrier model, we fitted the experimental data of the Hall effect and magnetoresistance simultaneously. The fitting curves for ${\varepsilon}=0${\%} are displayed in Figs.~\ref{fig:7}(c) and ~\ref{fig:7}(g). One can see that the experimental results are well reproduced. The obtained parameter sets, shown in Table~\ref{tab:T1}, are characterized by distinct two types of electron carriers. The first one shows similar parameters to those of the hole carrier, whereas the other has very small carrier density ${\neb}$ and high mobility ${\mueb}$. The present result is consistent with the previous transport study~\cite{Watson2015}. Taking into account the quantum oscillation~\cite{Terashima-SdH} and recent ARPES~\cite{Yi-ARPES} measurements, the first and second electron carriers likely arise from electron pockets at the Brillouin-zone corner consisting of $d_{xy}$ and $d_{yz}$ character for the former and mainly $d_{xy}$ character for the latter, while the hole carrier corresponds to a hole pocket at the zone center with $d_{xz}$ character.

Now let us move on to the strain dependence. Figures \ref{fig:7}(a)--\ref{fig:7}(h) show the fitting curves of the magneto-transport measurements on FeSe with four different degrees of biaxial strain. The magneto-transport data for even larger ${\varepsilon}$ are not shown because the Fermi-surface topology likely changes as mentioned above (see Appendix D). We summarize the fitting results of $n$ and ${\mu}$ as a function of ${\varepsilon}$ in Figs.~\ref{fig:7}(i) and ~\ref{fig:7}(j), respectively. The extracted parameters are also shown in Table~\ref{tab:T1}. All parameters exhibit a systematic strain dependence. ${\nh}$ and ${\nea}$  monotonically increase with decreasing ${\varepsilon}$. This agrees with the studies on thin films~\cite{Phan-ARPES,Nabeshima-Film-JJAP} which demonstrated that compressive strain enlarges the Fermi surface. On the other hand, the second electron carrier density ${\neb}$ shows a decrease. This behavior would arise from an upward shift of the $d_{xy}$ band due to compressive in-plane strain. Note that the shrinkage of the Fermi surface with decreasing ${\varepsilon}$ was not observed by the ARPES study~\cite{Phan-ARPES} probably because the electron pocket with $d_{xy}$ character is difficult to see due to its weak photoemission matrix elements~\cite{Yi-ARPES}. The increase in the net carrier density seems to be related with an increase in ${\Tc}$. The present result indicates that we definitely controlled the band structure of single-crystalline FeSe with application of biaxial strain.

Strain dependence is also observed for ${\mu}$. For all the three carriers, ${\mu}$ decreases with applying compressive strain. Since ${\mu}=e{\tau}/m^{*}$, where $m^{*}$ is the carrier effective mass and ${\tau}$ is the relaxation time, we have to consider the evolution of $m^{*}$ and ${\tau}$. As mentioned above, compressive strain induces the increase in the net carrier density, which is associated with an increase in band overlap between the hole and electron pockets~\cite{Phan-ARPES}. This probably stems from larger band width for a shorter lattice constant. Indeed, a lighter (heavier) effective mass $m^{*}$ is observed for compressive (tensile) in-plane strain~\cite{Phan-ARPES}. The change in the band structure also influences ${\tau}$ because a larger Fermi pocket gives rise to an expansion of momentum space available for scattering, leading to a decrease in ${\tau}$. The decrease in ${\mu}$ suggests that the effect of ${\tau}$ is more influential in the present case.

\section{Conclusion}

We succeeded in controlling the band structure of single-crystalline FeSe by applying in-plane biaxial strain. A wide range of biaxial strain from $-0.96\%$ to 0.23{\%} was applied for FeSe single crystals attached on the various substrates. The monotonic increase in $\Ts$ was observed as $\varepsilon$ is increased. The highest ${\Tc}$ was achieved for the most compressive strain, and ${\Tc}$ decreased with increasing ${\varepsilon}$. The analysis of the magneto-transport data using the compensated three-carrier model revealed the systematic strain dependence of $n$ and ${\mu}$ for all the three carriers. The present study highlights that biaxial strain is an effective tool to tailor physical properties of materials through the control of the band structure.

\begin{acknowledgments}
	This work was supported by The Murata Science Foundation and JSPS KAKENHI Grant number JP18K13500.
\end{acknowledgments}

\appendix
\section{Determination of absolute value of resistivity}

\begin{figure}
	\includegraphics[width=8cm]{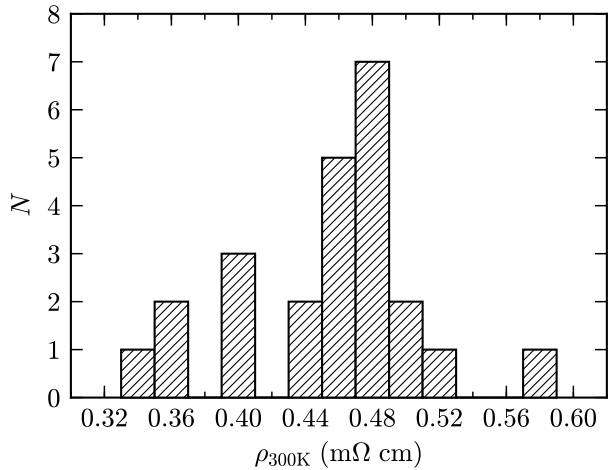}
	\caption{\label{fig:8}
		Histogram of the measured resistivity values for FeSe at 300 K.}
\end{figure}

To determine the absolute value of in-plane resistivity of FeSe single crystals at 300 K, we measured the resistivity for 27 samples. The histogram of the result is shown in Fig.~\ref{fig:8}. The average is calculated to be $0.45{\pm}0.05$ m${\Omega}$~cm, consistent with the previous study~\cite{Kasahara-Crossover}. Using the absolute value of the resistivity ${\rho}_{\rm 300K}$, we can estimate the thickness $t$ of the cleaved sample on the substrate by
\begin{align*}
t=\frac{{\rho}_{\rm 300K}}{R_{\rm 300K}}\frac{l}{w},
\end{align*}
where $R_{\rm 300K}$ is the resistance measured at 300 K, $l$ is the distance between the voltage terminals on the sample, and $w$ is the width of the sample.

\section{Analysis using compensated two-carrier model}

\begin{figure}
	\includegraphics[width=8.5cm]{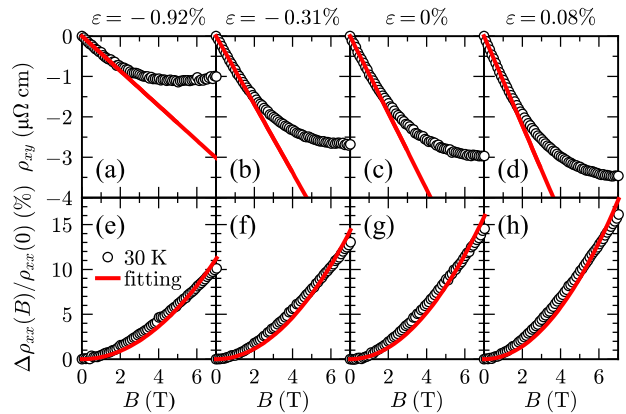}
	\caption{\label{fig:9}
		Magnetic-field dependence of (a)--(d) Hall resistivity and (e)--(h) magnetoresistance for strain-free and strained FeSe at 30 K. Fitting curves obtained using the compensated two-carrier model are also shown.}
\end{figure}

In a two-carrier model with hole and electron carriers, the Hall resistivity and magnetoresistance are expressed as follows:
\begin{align*}
{\rxy}(B) = \frac{1}{e} \frac{(\nh \muh^2 - \nd \mue^2) + \muh^2 \mue^2 B^2 (\nh - \nd)}{(\nh \muh + \nd \mue)^2 + \muh^2 \mue^2 B^2(\nh - \nd)^2} B
\label{rxy}
\end{align*}
and
\begin{align*}
\frac{{\rxx}(B)-{\rxx}(0)}{{\rxx}(0)}&=\frac{n_{\rm h}n_{\rm e}{\mu}_{\rm h}{\mu}_{\rm e}({\mu}_{\rm h}+{\mu}_{\rm e})^{2}B^{2}}
{(n_{\rm h}{\mu}_{\rm h}+n_{\rm e}{\mu}_{\rm e})^{2}+(n_{\rm h}-n_{\rm e})^{2}{\mu}_{\rm h}^{2}{\mu}_{\rm e}^{2}B^{2}},
\end{align*}
where ${\nh}$ (${\nd}$) and ${\muh}$ (${\mue}$) are the carrier density and mobility for holes (electrons), respectively. In the case for the carrier-compensation condition, i.e., ${\nh}={\nd}=n$, we obtain
\begin{align*}
{\rxy}(B) &=\frac{1}{ne}\frac{{\mu}_{\rm h}^{2}-{\mu}_{\rm e}^{2}}{({\mu}_{\rm h}+{\mu}_{\rm e})^{2}}B, \\
\frac{{\rxx}(B)-{\rxx}(0)}{{\rxx}(0)} &={\mu}_{\rm h}{\mu}_{\rm e}B^{2}.
\end{align*}
Since ${\rxy}$ shows a $B$-linear dependence in this model, we cannot reproduce the experimental data showing concave curvature. Thus, we fitted ${\rxy}$ up to 1 T and ${\rxx}$ up to 7 T simultaneously. We put a constraint that the calculated resistivity in zero field agrees with the value obtained experimentally [$1/{\rho}=ne({\muh}+{\mue})$]. The fitting results are displayed in Fig.~\ref{fig:9}.

\begin{figure}
	\includegraphics[width=7.5cm]{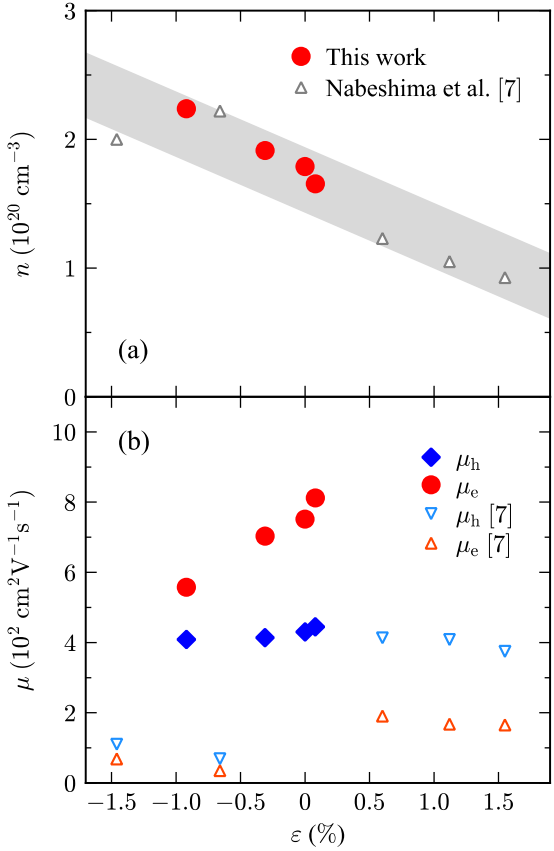}
	\caption{\label{fig:10}
		Strain dependence of (a) carrier density and (b) mobilities extracted from the magneto-transport data using the two-carrier model. Open triangle symbols represent the fitting result obtained from the thin-film study at 20 K \cite{Nabeshima-Film-JJAP}. The gray line is a guide to the eye.}
\end{figure}

In Fig.~\ref{fig:10}, we plot $n$ and ${\mu}$ obtained from the two-carrier fitting as a function of $\varepsilon$. The strain dependence is qualitatively in agreement with the result of the analysis using the three-carrier model [Figs.~\ref{fig:7}(i) and ~\ref{fig:7}(j)]. As the compressive strain becomes larger, the carrier density systematically increases. This is consistent with the previous thin-film results~\cite{Nabeshima-Film-JJAP} and can be explained by the increase of Fermi-surface volume~\cite{Phan-ARPES}. On the other hand, the strain dependence of carrier mobilities does not coincide with the thin-film study, which shows no systematic variation of the mobility with in-plane strain~\cite{Nabeshima-Film-JJAP}. Note that different from the present result, $B$-linear ${\rxy}$ was observed in the thin-film study, which can be well reproduced by the compensated two-carrier model. A difference is also seen in the sign of the Hall coefficient. In the present study, $\mue$ is larger than $\muh$, resulting in the negative Hall coefficient. In contrast, $\mue$ is smaller than $\muh$ for thin films, which gives the positive Hall coefficient. Electron carriers with high mobility seem to be absent for thin films of FeSe.

In addition to the non-linear $B$ dependence of ${\rxy}$, the magnetoresistance significantly deviates from the $B^{2}$ curve (see Fig.~\ref{fig:9}). To obtain more reasonable fitting results, we have to incorporate the contribution of the third carrier.

\section{Formalism of three-carrier model}

Three-carrier systems are characterized by carrier densities $n_j$ and mobilities $\mu_j$, where $j$ ($j$ = 1, 2, and 3) denotes the $j$th carrier component. Here, the sign of the carrier mobility is positive for holes and negative for electrons. The resistivity at zero magnetic field is
\begin{equation*}
	\rxx(0) = 1 / \sum_j |n_j e \mu_j|.
\end{equation*}
According to the three-carrier model~\cite{Kim1999} adopted in this work, the Hall resistivity and magnetoresistance are calculated by the following formulas:
\begin{align*}
	\rxy(B) &= \rxx(0) \left( \frac{\alpha + \beta_2 B^2 + \gamma_2 B^4}{1 + \beta_1 B^2 + \gamma_1 B^4} B \right), \\
	\frac{\rxx(B) - \rxx(0)}{\rxx(0)} &= \frac{\beta_3 B^2 + \gamma_3 B^4}{1 + \beta_1 B^2 + \gamma_1 B^4},
\end{align*}
where
\begin{align*}
	\alpha = &\; f_1 \mu_1 + f_2 \mu_2 + f_3 \mu_3, \\
	\beta_1 = &\; (f_1 \mu_2 + f_2 \mu_1)^2 + (f_2 \mu_3 + f_3 \mu_2)^2 + (f_1 \mu_3 + f_3 \mu_1)^2 \\
	              &\; + 2 (f_1 f_2 \mu_3^2 + f_2 f_3 \mu_1^2 + f_1 f_3 \mu_2^2), \\
	\beta_2 = &\; f_1 \mu_1 (\mu_2^2 + \mu_3^2) + f_2 \mu_2 (\mu_1^2 + \mu_3^2) + f_3 \mu_3 (\mu_1^2 + \mu_2^2), \\
	\beta_3 = &\; f_1 f_2 (\mu_1 - \mu_2)^2 + f_2 f_3 (\mu_2 - \mu_3)^2 + f_1 f_3 (\mu_1 - \mu_3)^2, \\
	\gamma_1 = &\; (f_1 \mu_2 \mu_3 + f_2 \mu_1 \mu_3 + f_3 \mu_1 \mu_2)^2, \\
	\gamma_2 = &\; (f_1 \mu_2 \mu_3 + f_2 \mu_1 \mu_3 + f_3 \mu_1 \mu_2) \mu_1 \mu_2 \mu_3, \\
	\gamma_3 = &\; f_1 f_2 (\mu_1 - \mu_2)^2 \mu_3^2 + f_1 f_3 (\mu_1 - \mu_3)^2 \mu_2^2 \\
	                 &\; + f_2 f_3 (\mu_2 - \mu_3)^2 \mu_1^2.
\end{align*}
In these expressions, $f_j$ is defined as
\begin{equation*}
	f_j = |n_j \mu_j| / \sum_j |n_j \mu_j|.
\end{equation*}

The three-carrier model is valid in the nonquantizing regime where Landau orbital quantization is negligible ($k_{\mathrm{B}} T \gg \hbar \omega_{\mathrm{c}}$) and within the relaxation-time approximation~\cite{Kim1993,Kim1998,Kim1999}. Strictly speaking, this model regards a conducting system associated with an energy band as a multicarrier system unless the relaxation time is energy independent. Each carrier component consists of charge carriers with the same mobility. An isotropic Fermi surface is assumed for each energy band so that the effective mass $m^*$ takes a constant value.

As described in the main text, we fitted the experimental data of the Hall resistivity and magnetoresistance simultaneously using the compensated three-carrier model. We considered one hole and two electron carrier components. Thus, the correspondence of each parameter is:
\begin{align*}
	\nh = n_1, \quad \nea = n_2, \quad \neb = n_3, \\
	\muh = \mu_1, \quad \muea = -\mu_2, \quad \mueb = -\mu_3.
\end{align*}

\section{Three-carrier fitting for $\varepsilon$ = 0.17\% and 0.23\%}

\begin{figure}[t]
	\includegraphics[width=7.5cm]{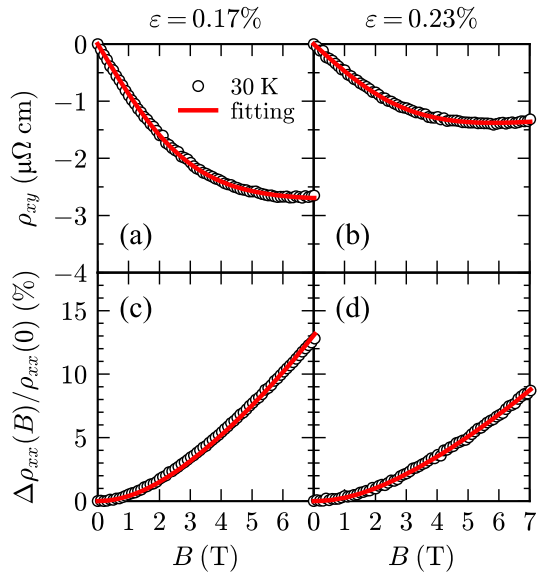}
	\caption{\label{fig:11}
		Magnetic-field dependence of (a)--(b) Hall resistivity $\rxy(B)$ and (c)--(d) magnetoresistance $\Delta \rho_{xx}(B) / \rho_{xx}(0)$ for strained FeSe with $\varepsilon$ = 0.17\% and 0.23\% at 30 K. Fitting curves obtained using the compensated three-carrier model are also shown.}
\end{figure}

\begingroup
\renewcommand{\arraystretch}{1.4}
\begin{table}
	\caption{\label{tab:T2}
		Carrier densities and mobilities extracted from the compensated three-carrier analysis for FeSe with $\varepsilon$ = 0.17\% and 0.23\% at 30 K.}
	\begin{center}
		\begin{tabular}{p{7.3em}p{3em}p{4.5em}p{4.5em}p{4.5em}} \hline \hline
			Parameters & \hfil ${\varepsilon}$ (\%) \hfil & \hfil hole \hfil & \hfil electron 1 \hfil & \hfil electron 2 \hfil \\ \hline
			$n$ ($10^{20}$ cm$^{-3}$) & \hfil 0.17 \hfil & \hfil 0.955 \hfil & \hfil 0.931 \hfil & \hfil 0.024 \hfil  \\
			${\mu}$  (cm$^{2}$V$^{-1}$s$^{-1}$)       &                    & \hfil 422  \hfil & \hfil  431 \hfil & \hfil 2063 \hfil  \\ \hline
			$n$ ($10^{20}$ cm$^{-3}$) & \hfil 0.23 \hfil & \hfil 1.521 \hfil & \hfil 1.495 \hfil & \hfil 0.026 \hfil  \\
			${\mu}$  (cm$^{2}$V$^{-1}$s$^{-1}$)       &                    & \hfil 341  \hfil & \hfil  341 \hfil & \hfil 1907 \hfil  \\ \hline \hline
		\end{tabular}
	\end{center}
\end{table}
\endgroup

Combined with the ARPES study~\cite{Phan-ARPES}, the present magneto-transport results indicate the presence of the Lifshitz transition between $\varepsilon$ = 0.08\% and 0.17\%. As shown in the inset of Fig.~\ref{fig:6}(b), a drastic change of the strain dependence was observed for both the Hall resistivity and the magnetoresistance. In principle, the three-carrier analysis can be carried out on the magneto-transport data for $\varepsilon$ = 0.17\% and 0.23\%, although comparison with the analysis for $\varepsilon \le 0.08\%$ should be carefully carried out.

The results of the compensated three-carrier analysis are displayed in Fig.~\ref{fig:11}. The obtained fitting parameters are shown in Table~\ref{tab:T2}. Although the experimental data are well reproduced by the fitting, we have to pay attention to how reasonable the parameters are.

\begin{figure}[H]
	\includegraphics[width=7.5cm]{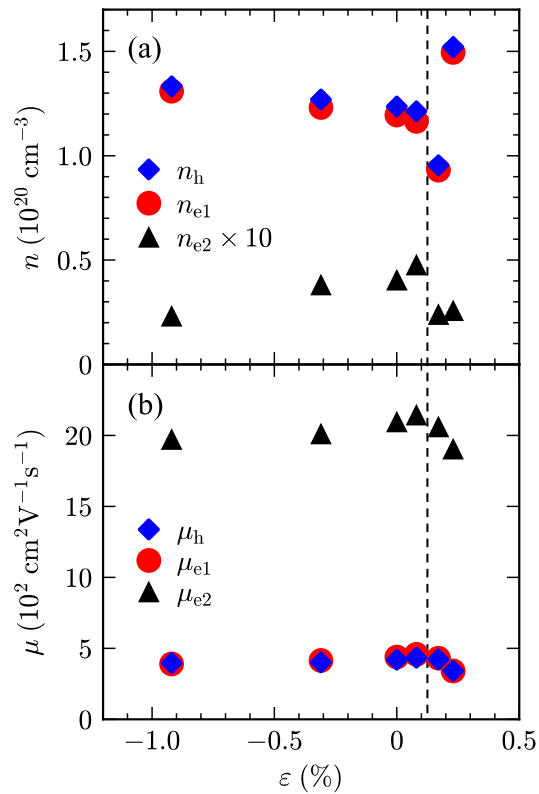}
	\caption{\label{fig:12}
		Strain dependence of (a) carrier densities (${\nh}$, ${\nea}$, and ${\neb}$) and (b) mobilities (${\muh}$, ${\muea}$, and ${\mueb}$) extracted by the three-carrier fitting shown in Fig.~\ref{fig:11}. A vertical broken line indicates the Lifshitz transition.}
\end{figure}

Figure~\ref{fig:12} shows the strain dependence of carrier densities and mobilities for one hole and two electron carriers. For $\varepsilon \le 0.08\%$, the parameters are the same as those plotted in Figs.~\ref{fig:7}(i) and 7(j). With increasing $\varepsilon$ across the Lifshitz transition, $\neb$ shows a drop, while no systematic variation is seen in $\nh$ and $\nea$. The mobilities for the three carriers decrease.

The large reduction of $\neb$ coincides with the insight extracted from the strain dependence of the magneto-transport data (Fig.~\ref{fig:6}). Since the second electron carriers possess high mobility, in spite of the low carrier density, the decrease in the contribution from this carrier component leads to a significant enhancement of the hole contribution as well as a decrease in averaged carrier mobility. The former makes $\rxy$ shift to the positive side, and the latter gives rise to smaller magnetoresistance.

\providecommand{\noopsort}[1]{}\providecommand{\singleletter}[1]{#1}%

\end{document}